\documentclass[useAMS,usenatbib]{mn2e}
\usepackage{epsfig}
\usepackage{amsmath}

\newcommand{\be}{\begin{equation}}
\newcommand{\beq}{\begin{equation}}
\newcommand{\ba}{\begin{eqnarray}}
\newcommand{\ee}{\end{equation}}
\newcommand{\eeq}{\end{equation}}
\newcommand{\ea}{\end{eqnarray}}

\def\lsim{~\rlap{$<$}{\lower 1.0ex\hbox{$\sim$}}}

\def\gsim{~\rlap{$>$}{\lower 1.0ex\hbox{$\sim$}}}

\title[High Redshift Black-Hole Seeds]{Photon Trapping Enables Super-Eddington Growth of Black-Hole Seeds in Galaxies at High Redshift}

\author[Wyithe \& Loeb]{J. Stuart B. Wyithe$^{1}$, Abraham Loeb$^{2}$\\$^1$
School of Physics, University of Melbourne, Parkville, Victoria,
Australia\\$^2$ Astronomy Department, Harvard University, 60 Garden
Street, Cambridge, MA 02138, USA\\Email: swyithe@physics.unimelb.edu.au; aloeb@cfa.harvard.edu}

\begin{document}

%\date{\today}
%\pagerange{\pageref{firstpage}--\pagebreak	eref{lastpage}} \pubyear{2006}

\maketitle

\label{firstpage}
\begin{abstract}

\noindent We identify a physical mechanism that would have resulted in rapid, obscured growth of seed super-massive black-holes in galaxies at $z\ga6$. Specifically, we find that the density at the centre of typical high redshift galaxies was at a level where the Bondi accretion rate implies a diffusion speed of photons that was slower than the gravitational infall velocity, resulting in photons being trapped within the accretion flow and advected into the black-hole. We show that there is a range of black-hole masses ($M_{\rm bh}\sim10^{3-5}$M$_\odot$) where the accretion flow traps radiation, corresponding to black-holes that were massive enough to generate a photon trapping accretion flow, but small enough that their Bondi radii did not exceed the isothermal scale height of self-gravitating gas. Under these conditions we find that the accretion reaches levels far in excess of the Eddington rate. A prediction of this scenario is that X-ray number counts of active galactic nuclei at $z\ga6$ would exhibit a cutoff at the low luminosities corresponding to black-hole masses below $\sim10^5$M$_\odot$. At low redshifts we find photon trapping to be unimportant because it could only occur in rare low spin halos, and would require black-hole masses in excess of expectations from the observed black-hole -- halo mass relation. The super-Eddington growth of $\sim10^{5}$M$_\odot$ seed black-holes at high redshift may have provided a natural acceleration towards the growth of super-massive black-holes at $z\sim6-7$, less than a billion years after the Big Bang.
\end{abstract}


\section{Introduction}

Exceptionally bright quasars with redshifts up to $z\sim7$ have been discovered \citep[e.g.][]{Fan2001,Fan2003,Mortlock2011}. These quasars are thought to be powered by the thin disk accretion of gas onto super-massive black-holes at the centres of galaxies. Their maximum (Eddington) luminosity depends on the mass of the black-hole, and the brighter quasars are inferred to have black-holes with masses of more than a few billion solar masses. Since their discovery \citep[][]{Fan2001,Fan2003}, the existence of such massive black-holes at $z\ga6$ has posed a challenge to models for their formation. This is because a $\sim10^9$M$_\odot$ black-hole accreting at the Eddington rate with a radiative efficiency of 10\% requires almost the full age of the Universe at $z\sim6$ to grow from a stellar mass seed. Many authors have therefore discussed solutions to this apparent mystery by including a significant build-up of mass through mergers \citep[][]{Haiman2001}, collapse of low-spin systems \citep[][]{Eisenstein1995}, and suppression of molecular line cooling via a large Lyman-Werner flux \citep[][]{Dijkstra2008}. Other authors have also attempted to explain the fast growth of black-holes at high redshifts based on super-massive stars \citep[][]{Loeb1994,Bromm2003} and more recently, quasi-stars \citep[][]{Volonteri2010,Begelman2010}. 

\citet[][]{Volonteri2005} discussed the possibility of super-Eddington accretion of black-hole seeds in high redshift galaxies. They pointed out that seed black-holes are located at the centres of isothermal disks where the conditions for quasi-spherical Bondi accretion should be prevalent. At high redshift these disk centres are sufficiently dense that the Bondi accretion rate greatly exceeds the Eddington rate. \citet[][]{Volonteri2005} point out that this super-Eddington accretion provides a route by which a large fraction of the mass $e$-foldings needed to grow a super-massive black-hole by redshift six could be accommodated within a small fraction of the age of the Universe. However the calculations in \citet[][]{Volonteri2005} ignored feedback effects like gas heating, which may raise the sound-speed, and hence lower the density and therefore the Bondi accretion rate. For example \citet[][]{Milosavljev2009} show that photoheating and radiation pressure from photoionization significantly reduce the steady-state accretion rate and potentially render a quasi-radial accretion flow unsteady and inefficient. They find that the time-averaged accretion rate is always a small fraction of the Bondi accretion rate. Thus, the very high accretion rates implied by the Bondi accretion in the centre of a high redshift isothermal disk might never be reached.  On the other hand, if the accretion rate is sufficiently high that the emergent photons are trapped within the accretion flow, then these feedback effects cannot operate \citep[][]{Begelman1979}, and so the accretion rate can reach arbitrarily high levels. 

In this paper we find that at sufficiently high redshift, the central densities of galaxies imply Bondi accretion rates that exceed the rate required to trap radiation and advect it into a black-hole \citep[][]{Begelman1979}. Thus, we find that there were periods in the growth of black-holes at high redshift where the growth was super-Eddington and feedback mechanisms could not halt the accretion flow.  Our goal is not to make a self-consistent model for both the transport of material from large galactic radii and central black-hole accretion. Rather, we note that there is a significant literature looking at the problem of super-Eddington accretion assuming large mass-delivery rates to the region of the black-hole, and identify the cosmological conditions that would provide sufficient accretion rates to allow this by trapping radiation.  We begin in \S~\ref{model} with a description of our simple model, before presenting our results in \S~\ref{results}. We finish with a discussion in \S~\ref{discussion}, and conclusions in \S~\ref{conclusion}.  In our numerical examples, we adopt the standard set of cosmological parameters \citep[][]{Komatsu2011}, with values of $\Omega_{\rm b}=0.04$, $\Omega_{\rm m}=0.24$ and $\Omega_\Lambda=0.76$ for the matter, baryon, and dark energy fractional density respectively, $h=0.73$, for the dimensionless Hubble constant, and $\sigma_8=0.82$.

\section{Model}

\label{model}
The basis of this paper is a comparison between the Bondi accretion rate, and the accretion rate required to trap photons within the accretion flow. We discuss these in turn.

\subsection{The Bondi accretion rate}

We begin with the expression for the Bondi accretion rate \citep[][]{Bondi1952}  onto a central black-hole of mass $M_{\rm bh}$ 
\begin{equation}
\label{bondi}
\dot{M}_{\rm Bondi} = 4\pi\rho_0 r_{\rm Bondi}^2 v_{\rm ff} = 4\pi n_0 \mu m_{\rm p} r_{\rm Bondi}^2 \sqrt{\frac{G M_{\rm bh}}{r_{\rm Bondi}}},
\end{equation}
where $v_{\rm ff}$ is the free-fall time at the Bondi radius
\begin{equation}
r_{\rm Bondi} = \frac{G M_{\rm bh}}{c_{\rm s}^2},
\end{equation}
and $c_{\rm s}$ is the sound speed, which for an isothermal gas we assume corresponds to a temperature of $10^4$K.
To evaluate $\dot{M}_{\rm Bondi}$ we specify the central number density of a self gravitating disk \citep[][]{Schaye2004} 
\begin{equation}
\label{n0}
n_0 = \frac{G M_{\rm disk}^2}{12 \pi c_{\rm s}^2 R_{\rm d}^4 \mu m_{\rm p}}.  
\end{equation}
Here $R_{\rm d}$ is the characteristic radius of an exponential disk of surface density profile
\begin{equation}
\Sigma(r) = \Sigma_0 \exp(-r/R_{\rm d}),
\label{eq:expdensprof}
\end{equation}
with $\Sigma_0 = M_{\rm disk}/ 2\pi R_{\rm d}^2$, and the disk scale length is given by
\begin{equation}
R_{\rm d} = {\lambda \over \sqrt{2}} r_{\rm vir},
\end{equation}
where $\lambda$ is the dimensionless spin parameter of the halo.
In equation~(\ref{n0}), $M_{\rm disk}=m_{\rm d}M_{\rm halo}$ is the disk mass, $m_{\rm p}$ is the mass of hydrogen, and $\mu=1.22$ is the molecular weight of primordial neutral gas. At the high redshifts of interest, most of the virialized galactic gas is expected to cool rapidly and assemble into the disk. We therefore assume $m_{\rm d}=0.17$. The corresponding mass density is $\rho_{\rm 0} = m_{\rm p} n_0$. The virial radius of a halo with mass $M_{\rm halo}$ is given by the expression
\begin{equation}
\label{eps}
\nonumber r_{\rm vir}= 0.784 h^{-1}\,\mbox{kpc} \left(\frac{M_{\rm halo}}{10^{8}M_{\odot}h}\right)^{\frac{1}{3}}
[\zeta(z)]^{-\frac{1}{3}}\left(\frac{1+z}{10}\right)^{-1},
\end{equation}
where $\zeta(z)$ is close to unity and
defined as $\zeta\equiv [(\Omega_{\rm m}/\Omega_{\rm m}^z)(\Delta_c/18\pi^2)]$,
$\Omega_{\rm m}^z \equiv [1+(\Omega_\Lambda/\Omega_{\rm m})(1+z)^{-3}]^{-1}$,
$\Delta_c=18\pi^2+82d-39d^2$, and $d=\Omega_{\rm m}^z-1$ \citep[see equations~22--25 in][for more details]{Barkana2001}. From equations~(\ref{n0}) and (\ref{eps}) we see that the central density $n_0$ and hence the Bondi accretion rate $\dot{M}_{\rm Bondi}$ scales as $(1+z)^4$, and that as a result accretion rates are expected to be much larger at high redshift.  

\begin{figure*}
\begin{center}
\vspace{3mm}
\includegraphics[width=17.5cm]{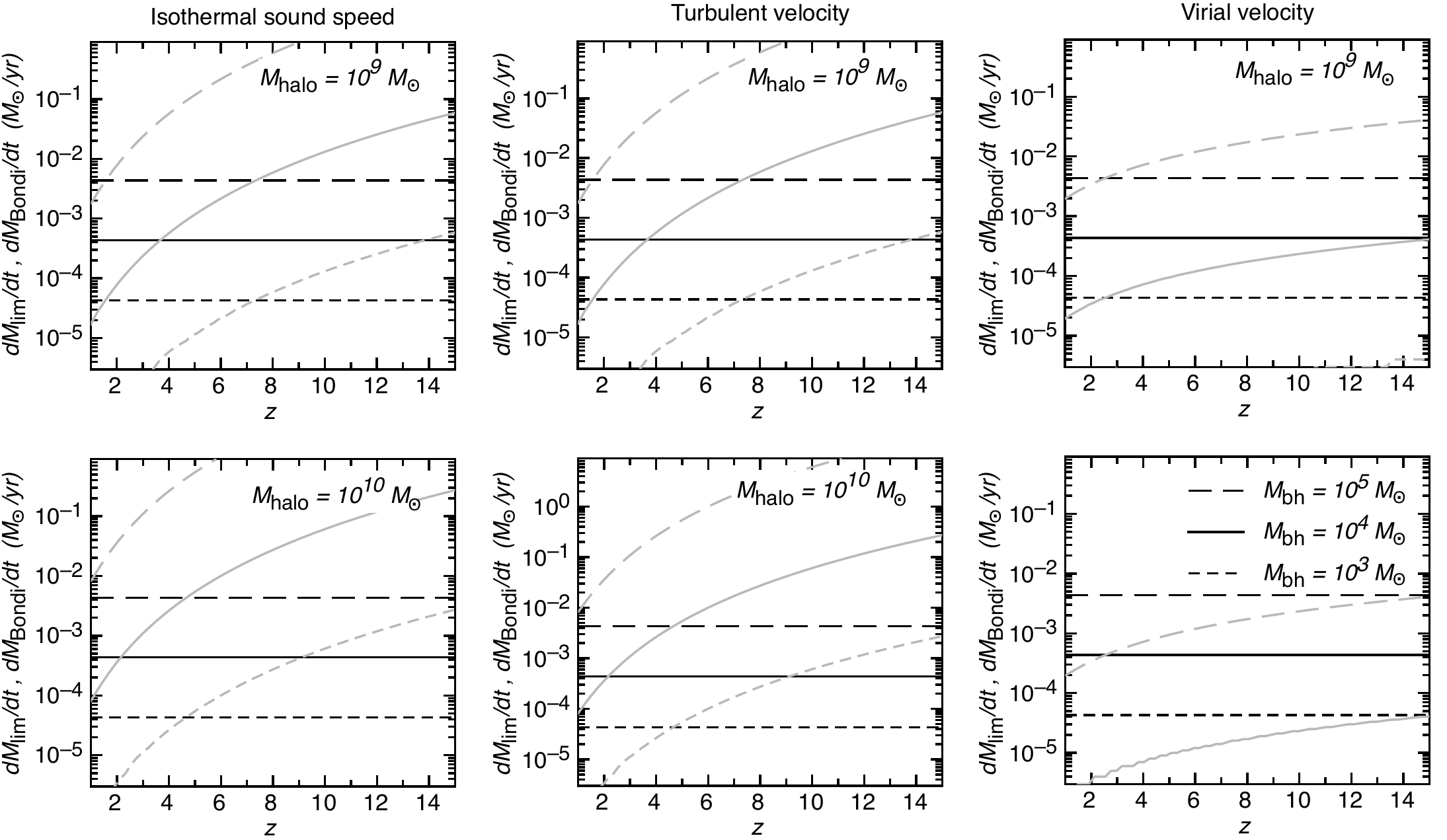}
\caption{\label{fig1} The grey curves show the Bondi accretion rate as a function of redshift. The dark lines show the critical accretion rate for which the photon diffusion speed is smaller than the gravitational free-fall speed. In cases where the Bondi accretion rate is larger than the critical accretion rate, photons are trapped and the AGN is obscured. Three cases are considered for the vertical structure of the disk, which is assumed to be set by the sound speed of the gas ({\em Left Panels}), the turbulent velocity associated with a $Q=1$ disk ({\em Central Panels}), and the virial velocity ({\em Right Panels}) respectively. In each case the {\em Upper} and {\em Lower} panels correspond to cases with halo masses of $M=10^{9}$M$_\odot$ and  $M=10^{10}$M$_\odot$. Three curves in each case correspond to black-hole masses of $10^3$M$_\odot$, $10^4$M$_\odot$ and $10^5$M$_\odot$. The cross-section was assumed to be $F_{\rm sig}=100$ times larger than the Thomson cross-section, the minimum radius was $F_{\rm min}=10^3$ times $r_{\rm g}$, and the spin parameter $\lambda=0.05$.}
\end{center}
\end{figure*}

Dormant central black-holes are ubiquitous in local galaxies \citep[][]{Magorrian1998}. The
masses of these super-massive black-holes scale with physical properties of their hosts
\cite[e.g.][]{Magorrian1998,Merritt2001,Tremaine2002}.  However at high redshift the relations observed in the local Universe may not be in place. We therefore do not impose a model for the relation between black-hole and halo mass in this paper, and instead explore a range of values. Indeed, our results indicate that feedback, which is thought to drive the black-hole -- halo-mass relation relation, would not be effective at early times. The grey curves in Figure~\ref{fig1} show the Bondi accretion rate as a function of redshift for different values of halo and black-hole mass. Here we assume $\lambda=0.05$ corresponding to the mean spin parameter for dark-matter halos \citep[][]{Mo1998}.

\subsection{Photon trapping by the accretion flow}

If the diffusion velocity of photons at a radius $r$ is smaller than the free-fall velocity of the material at radius $r$, then photons become trapped in the accretion flow \citep[][]{Begelman1979}. In such cases, the black-hole would be obscured. In this section we estimate the accretion rate required to achieve this photon trapping at radius $r$. 

The free-fall time from radius $r$ to a smaller radius $r_{\rm min}$ is 
\begin{equation}
\label{tff}
t_{\rm ff}\sim \frac{(r-r_{\rm min})}{v_{\rm ff}},
\end{equation}
where $v_{\rm ff}\sim \sqrt{G M_{\rm bh}/r}$ within the Bondi radius. This should be compared with the diffusion time of 
\begin{equation}
\label{tdiff}
t_{\rm diff}\sim \tau \frac{(r-r_{\rm min})}{c}.
\end{equation}
Here the optical depth is given by 
\begin{equation}
\tau \sim \bar{\rho} \frac{F_{\rm sig}\sigma_{\rm T}}{m_{\rm p}}(r-r_{\rm min}),
\end{equation}
where $\sigma_{\rm T}$ is the Thomson cross-section, $F_{\rm sig}$ is the ratio of the scattering cross-section of the accreting gas to the emergent radiation in units of $\sigma_{\rm T}$, and $\bar{\rho}$ is the line-of-sight averaged density between radii $r_{\rm min}$ and $r$. The density within the Bondi radius scales as $\rho(r)=\rho_0 (r/r_{\rm Bondi})^{-1.5}$, yielding
\begin{eqnarray}
\nonumber
\bar{\rho}&=&2\rho_0 \frac{r_{\rm Bondi}}{r-r_{\rm min}}  [(\frac{r_{\rm min}}{r_{\rm Bondi}})^{-0.5}  - (\frac{r}{r_{\rm Bondi}})^{-0.5}]\\
\nonumber
&=&2\rho \frac{r}{r-r_{\rm min}}  [(\frac{r}{r_{\rm min}})^{0.5}  - 1]\hspace{5mm}\mbox{for}\hspace{5mm}r<r_{\rm Bondi}.
\end{eqnarray}

The condition $t_{\rm diff}>t_{\rm ff}$ at radius $r$ is satisfied for accretion rates $\dot{M}>\dot{M}_{\rm lim}$ where
\begin{equation}
\dot{M}_{\rm lim}=\frac{\rho}{\bar{\rho}}\left(\frac{4\pi m_{\rm p} c}{F_{\rm sig}\sigma_{\rm T}}\right)\frac{r^2}{r-r_{\rm min}},
\end{equation}
in which we have used the relation $\dot{M}_{\rm Bondi} = 4\pi\rho r^2 v_{\rm ff}$. We find that the accretion rate required for photon trapping has a minimum value (i.e. $d\dot{M}_{\rm lim}/dr=0$) at a radius of $r=4r_{\rm min}$, yielding
 a minimum required accretion rate for photon trapping of
\begin{equation}
\label{lim}
\dot{M}_{\rm lim}=\left(\frac{8\pi m_{\rm p} c}{F_{\rm sig}\sigma_{\rm T}}\right)r_{\rm min}.
\end{equation}
Since the Eddington accretion rate at an efficiency $\epsilon$ is $\dot{M}_{\rm Edd}=(4\pi G M_{\rm bh} m_{\rm p})/(c\sigma_{\rm T}\epsilon)$, we find 
\begin{equation}
\label{lim2}
\dot{M}_{\rm lim}=2\left(\frac{\epsilon}{F_{\rm sig}}\right) \left(\frac{r_{\rm min}}{r_{\rm g}}\right) \dot{M}_{\rm Edd},
\end{equation} 
where $r_{\rm g}=GM_{\rm bh}/c^2$. 

The limiting accretion rate depends on the emergent radiation spectrum. As described below in \S~\ref{xray}, we find that the condition for photon trapping is more readily achieved for X-ray photons than for optical/UV photons. We therefore frame our discussion around trapping of optical/UV photons. In the case of optical/UV photons the cross-section would be dominated by dust at radii larger than a sublimation radius $r_{\rm min} = r_{\rm sub} $, beyond which the opacity of the gas to optical/UV photons could be larger than the Thomson opacity by as much as two or three orders of magnitude due to the presence of dust \citep[][]{Laor1993}. The existence of dust can be questioned for high-redshift galaxies \citep[e.g.][]{Bouwens2010d}, however high metallicities are inferred from the broad emission lines of all quasars out to $z=7.1$ and so metal enrichment (due to star formation) is known to precede the growth of the black-hole in the galactic nuclei of interest \citep[][]{Hamann2010}. On the other hand, the diffusion time may be lessened if the gas is clumpy, corresponding to a lower effective opacity. We therefore take a typical value for the cross-section that is $F_{\rm sig}=100$ times larger than the Thomson cross-section (additional cases are presented below in \S~\ref{structure}). Defining $r_{\rm min}=F_{\rm min} r_{\rm g}$ we assume a typical value of $F_{\rm min}=10^3$ to describe the sublimation radius \citep[][]{Netzer1993}. 

Limiting accretion rates corresponding to these default values are plotted in the left hand panels of Figure~\ref{fig1} (dark lines) for the case where the vertical structure of the disk is set by the sound speed of the gas. The {\em Upper} and {\em Lower} panels correspond to cases of halo masses of $M_{\rm halo}=10^{9}$M$_\odot$ and  $M_{\rm halo}=10^{10}$M$_\odot$. The larger value corresponds to the lower end of the inferred halo masses of the Lyman-break population at $z\ga6$ \citep[][]{trenti2010}. 

\subsection{Disk structure}
 \label{structure}

Before proceeding we first note that we have so far assumed the thickness of the gaseous disk at the centre of the galaxy to be set by the sound speed of gas at $10^4$K. However from equation~(\ref{ratio}) we see a strong dependence on the value for the effective sound speed. For example, in turbulent disks, the turbulent velocity replaces the isothermal velocity in determining the effective sound speed. Recently, \citet[][]{Genzel2010} inferred a Toomre's $Q$ value of $Q=1$ in ULIRGS, implying a value for the turbulent velocity of $c_{\rm T}\sim G \Sigma/\Omega=\sqrt{G\Sigma_0 r/\pi}$, where $v^2 = G\Sigma_0 \pi r^2 / r $,  yielding $\Omega = v/r = \sqrt{G\Sigma_0\pi/r}$. Evaluating at the Bondi radius,  we get
\begin{equation}
c_{\rm T} = \left(\frac{G^2\Sigma_0 M_{\rm bh}}{\pi}\right)^{0.25}.
\end{equation}
This value of $c_{\rm T}$ is the maximum value possible for a disk at large radius (as a higher $c_{\rm T}$ would imply an unphysical disk with $h>r$). Therefore, for a $Q=1$ disk the turbulent velocity should decrease towards small $r$. At sufficiently small radii, $c_{\rm T}<c_{\rm s}$, implying that the minimum thickness of the disk at its centre is set by the sound speed in the gas. Thus, although the turbulent velocity can be significant at large radii in the disk, we find that when evaluated at the Bondi radius, the turbulent velocity is smaller than the sound speed unless the halo is very massive and at high redshift ($\ga10^{11}$ at $z>4$). 

There may also be sources of energy injection that heat the gas to temperatures that correspond to velocities much larger than $c_{\rm T}$ (or $c_{\rm s}$), although in order to remain bound the velocity of the gas must be smaller than the virial velocity of the halo. We therefore show two cases in addition to $c_{\rm s}\sim10$km/s, namely $c_{\rm T}$ and $c_{\rm v}\sim f v_{\rm vir}$ where $f\sim0.5$ in order to bracket the range of interest, with corresponding limiting accretion rates plotted in the central and right hand panels of Figure~\ref{fig1} (dark lines). The results are almost unchanged where $c_{\rm T}$ is used to set the disk height. This is because we find that for the halo masses, black-hole masses and redshifts of interest, the sound speed is smaller than the turbulent velocity when evaluated at the Bondi-radius. However if the disk height is set by the maximum velocity $v_{\rm c}$, then larger black-holes and smaller halos are needed in order for the Bondi accretion rate to exceed the limiting rate, and at $z\sim6$ black-hole accretion would not be obscured in this case. In the remainder of this paper we restrict our attention to the case of an isothermal disk.

\section{Photon Trapping}
\label{results}

In cases where the Bondi accretion rate is larger than the critical accretion rate, rest frame optical/UV photons are trapped and the active galactic nucleus (AGN) can be obscured. Thus the cross-over of the limiting and Bondi accretion rate curves in Figure~\ref{fig1} represents the redshift beyond which accretion traps radiation for an AGN powered by the particular black-hole masses shown. If the vertical disk structure is set by the sound speed, we find that a $10^3$M$_\odot$ black-hole in a $10^{10}$M$_\odot$ halo will result in trapped radiation at $z>4$. In this section we discuss the range of halo and black-hole masses that result in photon trapping.

\begin{figure*}
\begin{center}
\includegraphics[width=17.5cm]{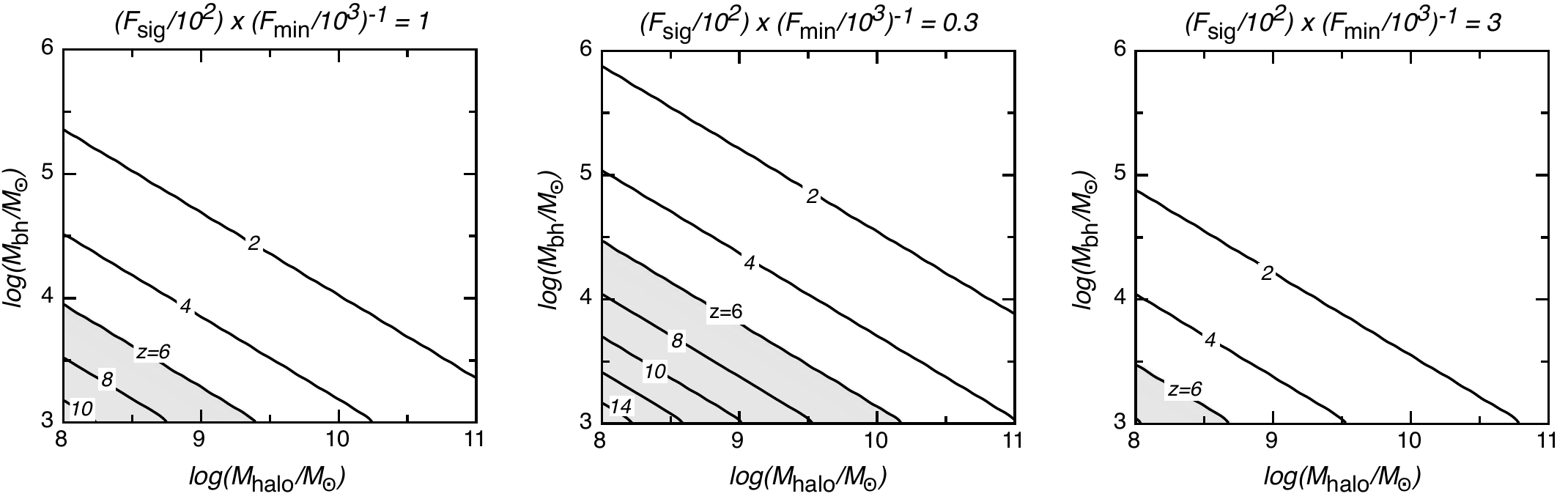}
\caption{\label{fig2}  Contours of the redshift at which the Bondi accretion rate becomes larger than the critical accretion rate as a function of black-hole and halo mass. We show contours for cases where the vertical structure of the disk is assumed to be set by the sound speed of the gas. In the {\em Left}, {\em Central} and {\em Right} panels we show contours for parameter combinations $(F_{\rm sig}/10^2)(F_{\rm min}/10^3)^{-1}=1$, $(F_{\rm sig}/10^2)(F_{\rm min}/10^3)^{-1}=0.3$ and $(F_{\rm sig}/10^2)(F_{\rm min}/10^3)^{-1}=3$. For illustration, the grey regions show the portion of parameter space at $z\sim6$ that does not result in photon trapping accretion flows. The example assumes $\lambda=0.05$ and $m_{\rm d}=0.17$.}
\end{center}
\end{figure*}

\subsection{Scaling relations}

The conditions for the halo mass, black-hole mass, and redshift that conspire to provide accretion rates that trap the optical/UV radiation can be obtained by combining equations~(\ref{bondi}) and (\ref{lim}). Evaluating we find
\begin{eqnarray}
\label{ratio}
\nonumber
&&\hspace{-7mm}\frac{\dot{M}_{\rm bondi}}{\dot{M}_{\rm lim}} \sim  37.5\left(\frac{M_{\rm bh}}{10^5\mbox{M}_\odot}\right) \left(\frac{M_{\rm halo}}{10^{10}\mbox{M}_\odot}\right)^{2/3}    \left(\frac{1+z}{7}\right)^4  \left(\frac{m_{\rm d}}{0.17}\right)^2  \\
&&\hspace{2mm} \times \left(\frac{F_{\rm min}}{10^3}\right)^{-1} \left(\frac{c_{\rm s}}{10\,\mbox{km/s}}\right)^{-5}\left(\frac{F_{\rm sig}}{100}\right)\left(\frac{\lambda}{0.05}\right)^{-4}.
\end{eqnarray}
This expression makes explicit the dependencies that lead to photons being more easily trapped (i.e. $\dot{M}_{\rm bondi}/\dot{M}_{\rm lim}>1$), namely larger black-holes in larger halos, and at higher redshift. We note that the condition for photon trapping has a much steeper dependence on redshift than obscuration via dust absorption, which scales as $(1+z)^2$.

This critical redshift at which the Bondi accretion rate exceeds the critical rate for photon trapping is shown graphically in Figure~\ref{fig2}. Here we  plot contours of the redshift at which the Bondi accretion rate becomes larger than the critical accretion rate, as a function of halo and black-hole mass. This figure shows the mass combinations that give a cross over from AGN in which radiation can escape to those in which the radiation is trapped. Based on equation~(\ref{ratio}) we plot contours for the parameter combinations $(F_{\rm sig}/10^2)(F_{\rm min}/10^3)^{-1}=1$, $(F_{\rm sig}/10^2)(F_{\rm min}/10^3)^{-1}=0.3$ and $(F_{\rm sig}/10^2)(F_{\rm min}/10^3)^{-1}=3$ in the {\em Left}, {\em Central} and {\em Right} panels of Figure~\ref{fig2}. Smaller (larger) values of this combination lead to larger (smaller) inflow rates being needed to trap the radiation, and so larger (smaller) black-holes are needed to trap the radiation at fixed halo mass and redshift (see equation~\ref{ratio}). 

From equation~(\ref{lim2}), the relation between the limiting rate for photon trapping and the Eddington rate can be illustrated  via the condition 
\begin{equation}
\label{ratio2}
\frac{\dot{M}_{\rm lim}}{\dot{M}_{\rm Edd}} \sim 2 \left(\frac{\epsilon}{0.1}\right)\left(\frac{F_{\rm sig}}{10^2}\right)^{-1} \left(\frac{F_{\rm min}}{10^3}\right).
\end{equation}
An accretion rate that leads to photon trapping is likely to be in excess of the Eddington rate, and hence associated with a very rapid build-up of black-hole mass.

\subsection{Trapping of X-rays}
\label{xray}

Up until now we have found that the critical rate at which the accretion flow traps optical/UV photons can be reached in the centres of high redshift galaxies owing to gas with large opacity (i.e. $F_{\rm sig}\gg1$) beyond a radius $F_{\rm min}r_{\rm g}$. The results of \citet[][]{Laor1993} indicate that while $F_{\rm sig}\gg1$ is expected for UV photons, the X-ray component of the spectrum will see an opacity to the inflowing gas that is set by the Thomson cross-section (i.e. $F_{\rm sig,X}=1$). Moreover, if the X-ray photon component of the spectrum were not trapped, it could halt the accretion flow via radiation pressure if it exceeded the Eddington rate by itself, thus preventing trapping of the optical/UV radiation. On the other hand, X-rays see this Thomson opacity at radii much smaller than the sublimation radius, with $F_{\rm min,X}\sim10$. 

To illustrate the importance of X-rays in this context we modify equation~(\ref{ratio2}) describing the relation between the limiting rate for optical/UV photon trapping and the Eddington rate that is specific to the X-ray portion of the spectrum ($\dot{M}_{\rm Edd,X}$). We define $F_{\rm X}$ to be the fraction of the luminosity in X-rays, which based on the spectrum in \citet[][]{Elvis1994} takes a value of $F_{\rm X}\sim10\%$. We therefore find 
\begin{equation}
\label{ratio3}
\frac{\dot{M}_{\rm lim}}{\dot{M}_{\rm Edd,X}} \sim 0.2 \left(\frac{\epsilon}{0.1}\right)\left(\frac{F_{\rm sig}}{10^2}\right)^{-1} \left(\frac{F_{\rm min}}{10^3}\right)\left(\frac{F_{\rm X}}{0.1}\right).
\end{equation}
Here $F_{\rm sig}\gg1$ and $F_{\rm min}\gg1$ correspond to the values for UV photon trapping. Equation~(\ref{ratio3}) shows that in order for X-rays to exceed the Eddington limit, the parameters corresponding to optical/UV trapping need to be $(F_{\rm sig}/10^2)(F_{\rm min}/10^3)^{-1}\ga5$. At these large accretion rates we find that when optical/UV photons are trapped, so too are the X-rays unless the minimum radius at which the X-rays encounter opacity ($r_{\rm min,X}=F_{\rm min,X}\,r_{\rm g}$) is described by a value of $F_{\rm min,X}>50$. To see this we note that the ratio of the Bondi accretion rate to the rate needed to trap X-rays is 
\begin{equation}
\frac{\dot{M}_{\rm Bondi}}{\dot{M}_{\rm lim,X}}=\left(\frac{F_{\rm min}}{F_{\rm min,X}}\right)\left(\frac{1}{F_{\rm sig}}\right)\frac{\dot{M}_{\rm Bondi}}{\dot{M}_{\rm lim}}.
\end{equation}
Since we expect $F_{\rm min,X}\sim6$, corresponding to the innermost stable circular orbit, we expect that an accretion flow which traps the optical/UV radiation will also trap the X-rays. As a result we do not consider the effect of X-rays on the accretion flow for the remainer of this paper.

\begin{figure*}
\begin{center}
\includegraphics[width=17.5cm]{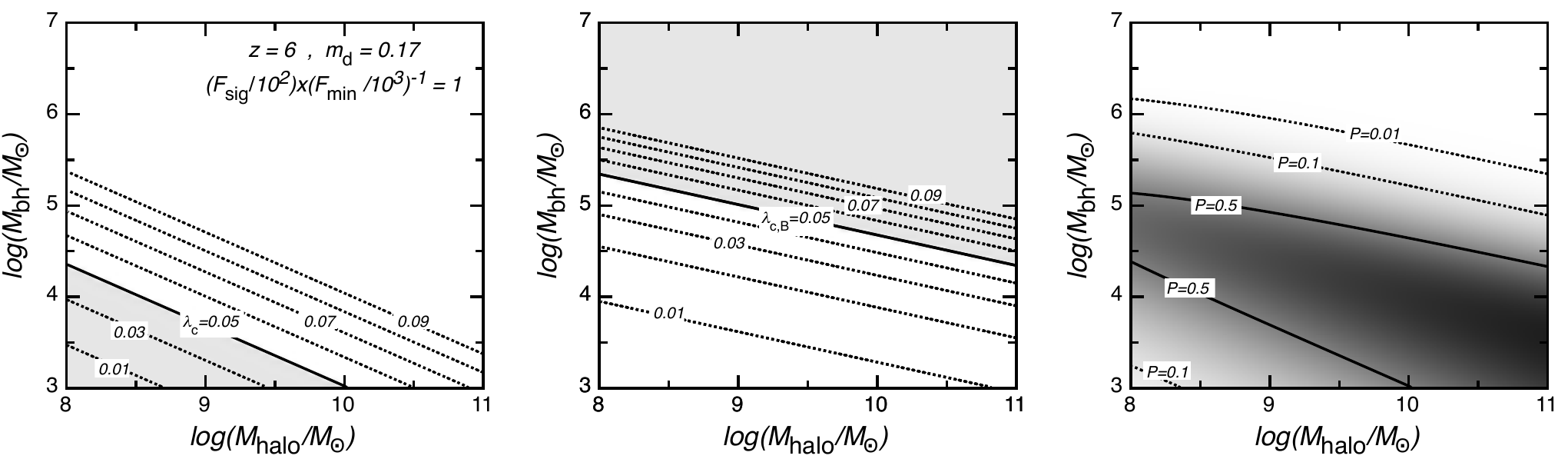}
\caption{\label{fig3_new} Regions of black-hole -- halo mass parameter space which result in photon trapping. {\em Left:} Contours of $\lambda_{\rm c}$ as a function of black-hole and halo mass. The grey shading shows the region of parameter space which does not result in photon trapping for the mean disk, because the critical spin parameter is smaller than $\lambda=0.05$. {\em Center:} Contours of $\lambda_{\rm c,B}$ as a function of black-hole and halo mass. The grey shading shows the region of parameter space which does not result in photon trapping for the mean disk, because the Bondi radius is larger than the scale height at the disk centre. {\em Right:} Contours of the probability that accretion will result in photon trapping, corresponding to the spin parameter $\lambda$ lying between the two critical values, i.e. $\lambda_{\rm c,B}<\lambda<\lambda_{\rm c}$. To calculate the probability we assume the distribution is Gaussian in the natural logarithm $\ln{\lambda}$, with variance $\sigma_\lambda=0.5$ and a mean at $\lambda=0.05$. Contours are shown for $P= 1\%$,  $P= 10\%$ and  $P= 50\%$. The examples assume $z=6$, $(F_{\rm sig}/10^2)(F_{\rm min}/10^3)^{-1}=1$, and $m_{\rm d}=0.17$. 
}
\end{center}
\end{figure*}

\subsection{Critical spin parameter}

Equation~(\ref{ratio}) shows that the trapping of photons is very sensitive to the value of the spin parameter $\lambda$ which governs the density of the galactic disk. In particular, photons may be more easily trapped within disks of low spin parameter. The assembly of dark matter halos leads to a distribution of spin parameters. To illustrate the parameter space we recast equation~(\ref{ratio}). Photon trapping at all wavelengths occurs for spin parameters $\lambda<\lambda_{\rm c}$ where
\begin{eqnarray}
\label{ratio4}
\nonumber
&&\hspace{-7mm}\lambda_{\rm c} \sim 0.12 \left(\frac{M_{\rm bh}}{10^5\mbox{M}_\odot}\right)^{1/4} \left(\frac{M_{\rm halo}}{10^{10}\mbox{M}_\odot}\right)^{1/6}    \left(\frac{1+z}{7}\right)  \left(\frac{m_{\rm d}}{0.17}\right)^{1/2}  \\
&&\hspace{7mm} \times \left(\frac{F_{\rm min}}{10^3}\right)^{-1/4} \left(\frac{c_{\rm s}}{10\,\mbox{km/s}}\right)^{-5/4} \left(\frac{F_{\rm sig}}{10^2}\right)^{1/4} ,
\end{eqnarray}

In the left panel of Figure~\ref{fig3_new} we plot contours of $\lambda_{\rm c}$ as a function of black-hole and halo mass at $z=6$. The example assumes $(F_{\rm sig}/10^2)(F_{\rm min}/10^3)^{-1}=1$, and $m_{\rm d}=0.17$. The mean spin parameter $\lambda=0.05$ is shown in black, and the grey shaded region illustrates the region of parameter space where trapping is not possible for a mean disk. Figure~\ref{fig3_new} shows that emergent photons from black-holes with $M_{\rm bh}\ga10^{3-4}$M$_\odot$ will be trapped at all wavelengths in the centres of high redshift galaxies ($\sim10^{8-10}$M$_\odot$) that are at the mean of the spin-parameter distribution .

\subsection{Breakout of the Bondi radius}
\label{breakout}

In the previous sections we have illustrated that the large densities expected at the centres of high redshift galaxies lead to conditions where accretion rates may be sufficient to trap photons within the accretion flow. These calculations are based on the density at the centre of a pressure supported self-gravitating disk. In this section, we point out that the calculation applies only to black-hole masses for which the Bondi radius is smaller than the scale-height at the disk centre. Moreover, we note that the trapping of radiation cannot be realised once the black-hole grows sufficiently that its Bondi radius exceeds the scale height. 

We again utilise the self-gravitating disk model. The scale height at the disk centre is 
\begin{equation}
\label{height}
z_0 = \frac{c_{\rm s}^2}{\pi G \Sigma_0}.
\end{equation}
This expression ignores the gravitational contribution from the black-hole which would serve to reduce the scale height, and is valid when $r_{\rm Bondi}< z_0$. 
Photon trapping and obscuration are only possible in this regime. 
We therefore calculate the ratio of Bondi radius to the central disk scale height as
\begin{equation}
\frac{r_{\rm Bondi}}{z_0} = \frac{G^2}{c_{\rm s}^4}\frac{M_{\rm bh}m_{\rm d}M}{\lambda^2R_{\rm vir}^2}.
\end{equation}
Putting in characteristic values we obtain
\begin{eqnarray}
\label{ratio5}
\nonumber\frac{r_{\rm Bondi}}{z_0} &=& 2.6 \left(\frac{M_{\rm bh}}{10^5\mbox{M}_\odot}\right) \left(\frac{M_{\rm halo}}{10^{10}\mbox{M}_\odot}\right)^{1/3}    \left(\frac{1+z}{7}\right)^2  \\
&&\hspace{9mm}\times\left(\frac{m_{\rm d}}{0.17}\right)  \left(\frac{c_{\rm s}}{10\,\mbox{km/s}}\right)^{-4}\left(\frac{\lambda}{0.05}\right)^{-2}.
\end{eqnarray}

Thus, within $M\sim10^{10}$M$_\odot$ halos at $z\sim6$, black-holes in excess of $M_{\rm bh}\sim10^5$M$_\odot$ have Bondi radii which are larger than the disk scale-height and so may not be obscured. We note that the ratio is very sensitive to the value of the sound speed, with a value larger than the fiducial $10\,$km$\,$s$^{-1}$ significantly reducing the ratio, allowing larger black-holes to accrete in the photon trapping mode. A small gas fraction also reduces the ratio. However the breakout of the Bondi radius implies that high mass black-holes such as those observed within the SDSS quasars \citep[e.g.][]{Fan2001,Fan2003} could not have their emergent radiation trapped.

To better understand the constraints imposed on black-hole masses where photon trapping can occur, we evaluate the critical value of spin parameter $\lambda_{\rm c,B}$ at which $r_{\rm Bondi}=z_0$
\begin{eqnarray}
\label{lambda_critB}
\nonumber
&&\hspace{-7mm}\lambda_{\rm c,B} \sim 0.08 \left(\frac{M_{\rm bh}}{10^5\mbox{M}_\odot}\right)^{\frac{1}{2}} \left(\frac{M_{\rm halo}}{10^{10}\mbox{M}_\odot}\right)^{\frac{1}{6}}    \left(\frac{1+z}{7}\right)  \left(\frac{m_{\rm d}}{0.17}\right)^{\frac{1}{2}}  \\
&&\hspace{10mm} \times  \left(\frac{c_{\rm s}}{10\,\mbox{km/s}}\right)^{1/2}.
\end{eqnarray}
In the central panel of Figure~\ref{fig3_new} we plot contours of $\lambda_{\rm c,B}$ as a function of black-hole and halo mass at $z=6$. The example again assumes $(F_{\rm sig}/10^2)(F_{\rm min}/10^3)^{-1}=1$, and $m_{\rm d}=0.17$. The mean spin parameter $\lambda=0.05$ is shown in black, and the grey shading illustrates the region of parameter space where photon trapping is not possible for the mean disk because the Bondi radius is larger than the scale-height. Figure~\ref{fig3_new} shows that black-holes with $M_{\rm bh}\ga10^{4.5}$M$_\odot$ within high redshift galaxies ($\sim10^{10}$M$_\odot$) having the mean of the spin-parameter cannot form a photon trapping accretion flow.

\subsection{When and where could photon trapping occur?}

The right panel of Figure~\ref{fig3_new} shows the probabilities that the spin parameter $\lambda$ lies between the two critical values needed for the conditions of $i)$ photon trapping, and $ii)$ a Bondi radius that is contained within the disk scale height (i.e. $\lambda_{\rm c,B}<\lambda<\lambda_{\rm c}$). To calculate this probability
\begin{equation}
P = \frac{1}{\sqrt{2\pi}\sigma_\lambda}\int_{\lambda_{\rm c,B}}^{\lambda_{\rm c}}  \exp{\left[-\frac{(\ln{\lambda}-\ln{\bar{\lambda}})^2}{2\sigma_\lambda^2}\right]}\,d\lambda
\end{equation}
 we assume the distribution of spin parameters is Gaussian in the natural logarithm $\ln{\lambda}$, with variance $\sigma_\lambda=0.5$ and a mean at $\bar{\lambda}=0.05$ \citep[][]{Mo1998}. The probability is $P=0$ if $\lambda_{\rm c}<\lambda_{\rm c,B}$, indicating a black-hole -- halo mass combination that cannot produce a photon trapping accretion flow. Contours are shown that represent black-hole -- halo mass combinations for which $P= 1\%$,  $P= 10\%$ and  $P= 50\%$ of disks would have densities that result in photon trapping. As before, the example assumes $(F_{\rm sig}/10^2)(F_{\rm min}/10^3)^{-1}=1$, and $m_{\rm d}=0.17$. The mean disk at $z\sim6$ has a central density that leads to photon trapping for black-hole masses up to $M_{\rm bh}\sim10^5$M$_\odot$ within halos up to $M_{\rm halo}\sim10^{9}$M$_\odot$. However $\sim10\%$ of disks are dense enough that photon trapping will occur for black-hole masses up to $M_{\rm bh}\sim10^5$M$_\odot$ within larger halos up to $M_{\rm halo}\sim10^{11}$M$_\odot$. Thus, we would expect photon trapping to be common in galaxies hosting $M_{\rm bh}\la10^5$M$_\odot$ black-holes at $z\sim6$. Moreover, since from equation~(\ref{ratio}) we see that this growth is super-Eddington, we find that photon trapping provides a mechanism by which rapid black-hole growth could proceed at high redshift, helping to explain how super-massive black-holes grow less than a billion years after the Big-Bang.

\begin{figure*}
\begin{center}
\includegraphics[width=17.5cm]{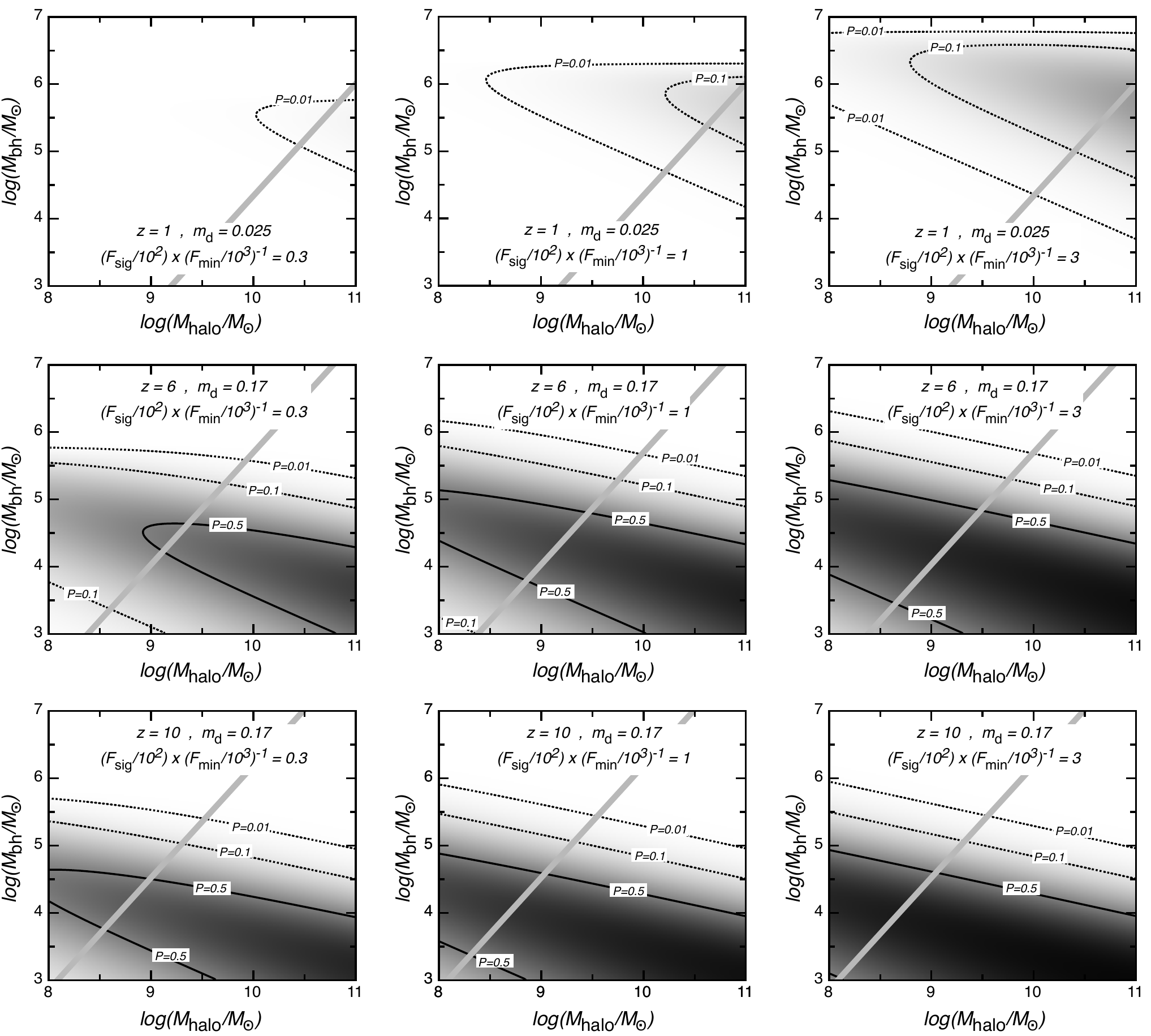}
\caption{\label{fig4_new}  Contours of the probability  in black-hole -- halo mass parameter space that accretion will result in photon trapping, corresponding to the spin parameter $\lambda$ lying between the two critical values, i.e. $\lambda_{\rm c,B}<\lambda<\lambda_{\rm c}$. Contours are shown for $P= 1\%$,  $P= 10\%$ and  $P= 50\%$. Examples are shown for three values of redshift, $z=1$, $z=6$ and $z=10$ ({\em Top} to {\em Bottom}). For $z=6$ and $z=10$ we choose $m_{\rm d}=0.17$. We assume $m_{\rm d}=0.025$ at $z\sim1$.  In each case we show examples with $(F_{\rm sig}/10^2)(F_{\rm min}/10^3)^{-1}=0.3$, $(F_{\rm sig}/10^2)(F_{\rm min}/10^3)^{-1}=1$ and $(F_{\rm sig}/10^2)(F_{\rm min}/10^3)^{-1}=3$ (from {\em Left} to {\em Right}). The thick grey curve is the predicted value for black-hole mass halo mass relation (equation~\ref{relation}).}
\end{center}
\end{figure*}

In Figure~\ref{fig4_new} we explore how the conclusions regarding the black-hole and halo mass-ranges that produce photon trapping accretion flows are effected by redshift and the parameter combination  $(F_{\rm sig}/10^2)(F_{\rm min}/10^3)^{-1}$. We choose three values of redshift, $z=1$, $z=6$ and $z=10$. For $z=6$ and $z=10$ we choose $m_{\rm d}=0.17$. However at later times we expect that gas is less plentiful and so assume $m_{\rm d}=0.025$ at $z\sim1$, corresponding to a comparison with low redshift disks \citep[][]{Mo1998}. These cases are shown in the upper central and lower rows respectively. In each case we show examples with $(F_{\rm sig}/10^2)(F_{\rm min}/10^3)^{-1}=0.3$, $(F_{\rm sig}/10^2)(F_{\rm min}/10^3)^{-1}=1$ and $(F_{\rm sig}/10^2)(F_{\rm min}/10^3)^{-1}=3$ ({\em Left}, {\em Central} and {\em Right} panels respectively). In all panels we show contours for the probabilities that the spin parameter $\lambda$ lies between the two critical values needed for the conditions of $i)$ photon trapping, and $ii)$ a Bondi radius that is contained within the disk scale height (i.e. $\lambda_{\rm c,B}<\lambda<\lambda_{\rm c}$). As before contours are shown that represent black-hole -- halo mass combinations for which $P= 1\%$,  $P= 10\%$ and  $P= 50\%$ of disks would have central densities that lead to photon trapping.

We find that larger values of $(F_{\rm sig}/10^2)(F_{\rm min}/10^3)^{-1}$ lead to more massive black-holes in smaller halos having their emergent radiation trapped. At redshift $z\sim6-10$ we find that black-holes with masses up to $M_{\rm bh}\sim10^5$M$_\odot$ have their radiation trapped in 10\%-50\% of cases. At lower redshift,  black-holes with masses of $M_{\rm bh}\sim10^{5.5}$M$_\odot$ could have radiation trapped in up to 10\% of cases, but only if the halo mass and  $(F_{\rm sig}/10^2)(F_{\rm min}/10^3)^{-1}$ are large. Thus, photon trapping is likely to be a phenomenon dominated by low mass ($M_{\rm bh}\la10^5$M$_\odot$) black-holes in high redshift ($z\ga6$) galaxies ($M_{\rm halo}\sim10^{10}$M$_\odot$).

\subsection{Comparison with $M_{\rm bh}-M_{\rm halo}$ models}

As noted in the introduction, the relations observed between black-hole mass and galaxy properties in the local Universe may not be in place at high redshift. For this reason we have not imposed a model for the relation between black-hole and halo mass in this paper, and instead have explored a range of values. However it is interesting to compare the range of black-hole masses found to be accreting in the photon trapping mode with expectations of the simple models relating black-hole and halo masses that have been successful in describing some of the properties of high redshift quasars. 

Motivated by local observations \citep[][]{Ferrarese2002}, we consider a model in which the central black-hole mass is correlated with the halo
circular velocity. This scenario is supported by the results of \citet[][]{Shields2003} who studied quasars out to $z\sim3$ and demonstrated that the
relation between black-hole masses and the stellar velocity dispersion does not
evolve with redshift. This is expected if the mass of the black-hole is
determined by the depth of the gravitational potential well in which it
resides, as would be the case if growth is regulated by feedback from
quasar outflows \cite[e.g.][]{Silk1998,Wyithe2003b}.  Expressing
the halo virial velocity, $v_c$, in terms of the halo mass, $M_{\rm halo}$, and
redshift, $z$, the redshift dependent relation between the super-massive black-hole and halo
masses may be written as
\begin{equation}
\label{relation}
M_{\rm bh} = \epsilon_{\rm bh} M_{\rm halo} \left(  \frac{M_{\rm halo}}{10^{10}\mbox{M}_\odot}\right)^{2/3}[\zeta(z)]^{5/6} \left(\frac{1+z}{7}\right)^{5/2}.
\end{equation}
The normalising constant in this relation has an
observed\footnote{ We have used the normalization derived by
\citep[][]{Ferrarese2002} under the simplifying assumption
that the virial velocity of the halo
represents its circular velocity.} value of $\epsilon_{\rm bh}\approx 10^{-4.3}$ based on calibration of equation~(\ref{relation}) at $z=0$ \citep[][]{Ferrarese2002}, where we take the underlying
assumption that the halo mass profile resembles a Singular Isothermal Sphere. In \citet[][]{Wyithe2005} this model was shown to be consistent with the clustering and luminosity function data of the 2dF quasar redshift survey \citep[][]{Croom2002}. 

Interestingly, the predicted value for super-massive black-hole masses in galaxy mass halos (as shown by the thick grey curve in Figure~\ref{fig4_new}) is comparable to the range at which we would expect photon trapping in galaxies with $z\ga6$. Thus, prior to obtaining masses where the self regulating mechanisms thought to be responsible for the relations between black-hole mass and halo properties take effect we would expect black-holes in high redshift galaxies to reach the level where super-Eddington accretion would become a natural part of their evolution. However at low redshift, Figure~\ref{fig4_new} shows that photon trapping requires black-hole masses that are greater than the observed black-hole -- halo mass relation. Thus, we do not expect expect photon trapping within low redshift galaxies.

\section{Discussion}
\label{discussion}

The findings in this paper point to some potentially important implications for the growth of high redshift super-massive black-holes. Recent simulations by \citet[][]{Li2011} show that self gravity overcomes radiative feedback and that accretion onto intermediate mass black-holes reaches the Eddington rate. However, at high enough accretion rates, a spherical inflow is not subject to the Eddington limit (though the emergent radiation is), and the Bondi accretion rate can take arbitrarily large values in high density galactic centres \citep[][]{Begelman1979}. The conditions for spherical accretion may only rarely be realised at the centres of high redshift galaxies. On the other hand, there is a class of slim disk models \citep[e.g.][]{Abramowicz1988,Chen1995,Ohsuga2005,Watarai2006,Ohsuga2011}, for which super-Eddington accretion flows are possible.  These models include a large viscosity, and so rapid transport of material through the disk within a small multiple of the free fall time \citep[e.g.][]{Watarai2006}, and are optically thick and advection dominated \citep[][]{Chen1995}. Photon trapping plays an important role within these disks, even those with complex flows, leading to very inefficient energy conversion and a luminosity that is independent of accretion rate \citep[e.g.][]{Ohsuga2005}. Simulations indicate that at sufficient densities, the mass accretion rate can reach hundreds of Eddington, although the photon luminosity does not \citep[e.g.][]{Ohsuga2011}. These super-Eddington accretion disks will generate outflows \citep[which cannot be generated in spherical accretion,][]{Begelman1979} that are likely to be channelled along the poles, and so will not suppress the accretion along the equatorial plane. The calculations presented in this paper provide the boundary condition for super-Eddington flows. These rapid accretion events may provide the seeds for super-massive black-hole growth.

A plausible scenario therefore includes two episodes of black-hole growth. Initially the accretion may have been spherical or via a slim disk with high viscosity, so that the Eddington rate was greatly exceeded by the accretion rate. We find that radiation would have been trapped  and advected into the central black-hole by the very large Bondi accretion rates at high redshift. This phase of growth would be obscured at both optical/UV and X-ray wavelengths. Once the Bondi radius became larger than the scale-height of the galaxy disk, the accretion rate dropped, and higher
angular momentum gas would have settled into a thin disk accretion mode, in which the
accretion time was much longer than the free-fall time, allowing the radiation to escape. For a pressure supported disk, we have found that the Bondi radius is expected to exceed the scale-height of the disk for black-holes in excess of $\sim10^{5}$M$_\odot$.
The luminous quasars discovered at $z\gsim6$ with black-hole masses in excess of $10^8$M$_\odot$ are therefore thought to be shining in this later mode, un-obscured in the optical and with accretion rates close to, but smaller than Eddington. However a portion of their prior growth, when the black-hole mass was $\sim10^{4-5}$M$_\odot$ would have been in the photon trapping mode.

Super-Eddington accretion rates arising from the large densities in the cores of high redshift galaxies were previously considered by \citet[][]{Volonteri2005}. As noted in the introduction, the calculations in \citet[][]{Volonteri2005} neglected feedback effects like gas heating, which may lower the the Bondi accretion rate \citep[e.g.][]{Milosavljev2009}.  In this paper we note that if the accretion rate is sufficiently high that the emergent photons are trapped within the accretion flow, then feedback effects cannot operate \citep[][]{Begelman1979}. Our model also makes two different assumptions which modify our conclusions relative to \citet[][]{Volonteri2005}. Firstly, \citet[][]{Volonteri2005} assume that once the gas is enriched, metal line cooling allows the gas to cool to temperatures much lower than $10^4$K, so that it fragments to form stars and the super-Eddington accretion episode is ended. This assumption was necessary in order that super-Eddington accretion not lead to black-hole densities in excess of those observed. However, we expect that fragmentation and associated star-formation will reheat the gas via radiative and supernova feedback so that the pressure support at the centre of the disk is maintained and accretion can continue. This scenario is supported by the observations of \citet[][]{Genzel2010} which imply significant turbulent velocities in high redshift galaxies. Rather than make an arbitrary assumption that growth stops when the size of the accretion disk grows by a factor of five, we instead assume that the super-Eddington photon trapping accretion mode would be regulated by the time when the Bondi radius exceeds the scale height of the pressure supported disk. We find that this condition prevents super-Eddington accretion onto high mass black-holes.

\subsection{Obscured accretion}

Our results have some relevance to the recent discussion surrounding obscured accretion in high redshift galaxies. While luminous optical quasars in the most massive halos ($\ga10^{12}$M$_\odot$) dominate the observations of high redshift super-massive black-holes, \citet[][]{Treister2011} recently presented evidence that most of the black-hole accretion at $z\ga6$ is actually optically obscured, and in galaxies below halo masses of $\sim10^{10-11}$M$_\odot$. Since high redshift Lyman-break galaxies are thought not to be dusty \cite[e.g.][]{Bouwens2010d}, our results might have provided a mechanism by which the AGN can be obscured even in the absence of a large dusty component. However our results do not support a photon trapping explanation for this result. Firstly, X-ray photons would be more strongly trapped by the accretion flow than the UV photons, indicating that we would not expect to observe X-rays without optical detection. In addition, we also find that photon trapping is only expected in a fraction of galaxies rather than in all galaxies as implied by \citet[][]{Treister2011}. 

While other mechanisms may lead to buried accretion, the findings of \citet[][]{Treister2011} have also been disputed by a number of authors \citep[][]{Fiore2011,Cowie2011,Willott2011}, who do not observe the same level of X-ray emission. These authors find limits on the average X-ray luminosity in the rest frame 0.5-2keV band of $L_{0.5-2}<4\times10^{41}\,$erg/s for $z\sim6.5$ dropout galaxies. This luminosity can be related to black-hole mass as
\begin{equation}
L_{0.5-2}\sim 3\times10^{41}\left(\frac{M_{\rm bh}}{10^5\mbox{M}_\odot}\right)\eta\,\mbox{erg}\,\mbox{s}^{-1},
\end{equation}
where $\eta=1$ is the fraction of the Eddington accretion rate, and we have assumed the spectral energy distribution of \citet[][]{Elvis1994}. Observed luminosities must have $\eta<1$ even if the accretion rate is super-Eddington \citep[][]{Begelman1979}. The observed luminosity limit therefore corresponds to observed black-hole masses of $M_{\rm bh}\la1.3\times 10^5$M$_\odot$. We do not find that black-holes with masses $M_{\rm bh}\ga10^5$M$_\odot$ produce photon trapping accretion flows, and so photon trapping does not explain the lack of observed X-ray sources among the $z\sim6.5$ dropouts \citep[][]{Fiore2011,Cowie2011,Willott2011}. Since the stacked observations in these studies are based on only $\sim10^2$ galaxies, this lack of detection could follow from the low-duty cycle of AGN \citep[which is likely a few percent,][]{Wyithe2003b}. However our results do suggest that 90\%-100\% of disks with black-holes below this mass would be in the photon trapping mode. Thus, we would expect that deeper and wider field X-ray observations using future X-ray observatories, to display a cut-off in the X-ray luminosity function at about $L_{0.5-2}\sim 3\times10^{41}$erg$\,$s$^{-1}$. Conversely, the discovery of X-ray AGN with luminosities an order of magnitude lower than current limits would therefore rule out photon trapping accretion as a mechanism for rapid growth of early black-holes. Finally, we note that since super-Eddington accretion at high redshift is obscured at both optical and X-ray wavelengths, rapid growth of seed black-holes could not provide a significant source of X-ray for reionization of the IGM \citep[][]{Volonteri2005}.

\subsection{Seed black-hole growth}

The photon trapping mode is  likely to be important for the rapid growth of seed super-massive black-holes with masses of $\sim10^{4-5}$M$_\odot$. Because the Bondi accretion rates in these high redshift galaxies could be  orders of magnitude larger than the Eddington rate, the photon trapping mechanism helps alleviate the difficulty of growing super-massive black-holes of more than a billion solar masses \citep[corresponding to the most distant known quasars,][]{Mortlock2011} within the first billion years of the Universe's age. This point was made in detail in \citet[][]{Volonteri2005}. To illustrate we note that accretion at the Eddington rate (with $\epsilon=0.1)$ leads to an $e$-folding time of $t=4\times10^7$ years. Assuming that a black-hole accretes with a duty cycle of unity, the number of e-folding times available by $z\sim7$ is therefore $\sim20$. This should be compared with the 20 $e$-folds needed to grow a $1$M$_\odot$ black-hole seed up to a mass of $10^9$M$_\odot$.  Thus there is only just enough time during the age of the Universe at $z\ga6$ for a stellar black-hole seed to grow to a super-massive black-hole. We suggest that the period of obscured growth in some galaxies would provide a path toward growing these super-massive black-holes.

\section{Conclusion}
\label{conclusion}

In this paper we have determined the cosmological regime in which photons produced through accretion onto a central black-hole are trapped by infalling material, so that $i)$ radiation feedback on the infall of gas outside the dust sublimation radius is suppressed, allowing accretion rates far in excess of the Eddington limit, $ii)$ AGN appear obscured, and $iii)$ the growth time of black-hole is short. Specifically we find that a large fraction of galaxies at $z\ga6$ with masses up to those of the observed Ly-break population (halo masses of $\sim10^{9-11}$M$_\odot$) exhibit Bondi-accretion rates onto $M_{\rm bh}\sim10^{3-5}$M$_\odot$ black-holes that are sufficiently high to trap the resulting rest frame optical/UV/X-ray radiation. The obscuration due to photon trapping is found only to occur for black-holes with masses up to $\sim10^5$M$_\odot$ because larger black-holes have Bondi radii that exceed the scale height of the disk from which they accrete gas, so that a photon trapping mechanism cannot operate. As a result, we find a natural distinction between obscured, photon trapping accretion onto $\sim10^5$M$_\odot$ black-holes in galaxies of halo mass $\la10^{10}$M$_\odot$, and the luminous accretion seen in the brightest quasars with black-hole masses of $\sim10^{8-9}$M$_\odot$ within halos of mass $\sim10^{11-12}$M$_\odot$. At lower redshift photon trapping requires black-holes that are larger than expected from the black-hole -- halo mass relation, and so is not expected to be observed. Our results indicate that super-Eddington accretion of mass to form seed black-holes of $\sim10^5$M$_\odot$ provided a mechanism by which super-massive black-holes were able to form prior to $z\sim6$. 

\vspace{5mm}

{\bf Acknowledgments} JSBW acknowledges
the support of the Australian Research Council.  AL was supported in
part by NSF grant AST-0907890 and NASA grants NNX08AL43G and
NNA09DB30A.

\newcommand{\noopsort}[1]{}

\bibliographystyle{mn2e}

\bibliography{text}

\label{lastpage}
\end{document}